\documentclass{llncs}

\begin{document}
\spnewtheorem{theor}[subsubsection]{Theorem}{\bfseries}{\rmfamily}
\spnewtheorem{defi}[subsubsection]{Definition}{\bfseries}{\rmfamily}

\title{
A New General-Purpose Method
to Multiply 3x3 Matrices
Using Only 23 Multiplications$^{*}$
}
\titlerunning{}
\toctitle{}

\author{
Nicolas T. Courtois\inst{1,3} \and
Gregory V. Bard\inst{2} \and
Daniel Hulme\inst{1,3}  
\vskip-3pt
\vskip-3pt
}

\institute{
$^{1}$ University College London, UK,\\
$^{2}$ University of Wisconsin Stout Campus, Menomonie, WI, USA\\
$^{3}$ NP-Complete Ltd, London, UK
}

\maketitle

\begin{abstract}
One of the most famous conjectures in computer algebra
is that matrix multiplication might be feasible in 
nearly quadratic time, \cite{KohnGroupBreaktroughFastMM}. 
The best known exponent is $2.376$,
due to Coppersmith and Winograd \cite{CPW}.

Many attempts to solve this problems in the literature work by solving,
fixed-size problems and then apply the solution recursively
\cite{Brent70,PanSpeed,Kaporin10K,IP6,BardApprox22}. 
This leads to pure combinatorial optimisation problems with fixed size.
These problems are unlikely to be solvable in polynomial time, see \cite{IP6,has}.

In 1976 Laderman published a method to multiply two 3x3 matrices using only 23 multiplications.
This result is non-commutative,
and therefore can be applied recursively to smaller sub-matrices.
In 35 years nobody was able to do better
and it remains an open problem if this can be done with 22 multiplications.

We proceed by solving the so called Brent equations \cite{Brent70}.
We have implemented a method to converting this very hard problem to a SAT problem,
and we have attempted to solve it,
with our portfolio of some 500 SAT solvers. 
With this new method
we were able to produce new solutions to the Laderman's problem.

We present a new fully general non-commutative solution with 23 multiplications
and show that this solution is new and
is {\bf NOT} equivalent to any previously known solution.
This result demonstrates that the space of solutions
to Laderman's problem is larger than expected,
and therefore it becomes now more plausible
that a solution with 22 multiplications exists.

If it exists, we might be able to find it soon
just by running our algorithms longer,
or due to further improvements
in the SAT solver algorithms.
\end{abstract}


\paragraph{}
$^{\mathbf{*}}$
This work was supported by the Technology Strategy Board
in the United Kingdom under Project No 9626-58525.

\newpage
\section{Introduction}

One of the most famous problems in computer algebra is
the problem of matrix multiplication (MM) of square and non-square matrices.

\subsection{Fast Matrix Multiplication}

For square matrices the naive algorithm is cubic 
and the best known theoretical exponent is $2.376$,
due to Coppersmith and Winograd \cite{CPW}. 
This exponent is quite low and 
it is conjectured that one should be able to 
do matrix multiplication 
in so called ``soft quadratic time'', 
with possibly some poly-logarithmic overheads, 
which could even be sub-exponential in the logarithm, 
\cite{KohnGroupBreaktroughFastMM}. 
This in fact would be nearly- linear in the size of the input (!). 


In 2005 a team of scientists
from Microsoft Research and two US universities
established a new method for finding such algorithms based on group theory,
and their best method so far gives an exponents of 2.41 \cite{KohnGroupBreaktroughFastMM},
very close to Coppersmith-Winograd result
and subject to further improvement.

It is also known that efficient algorithms for fast matrix multiplication
are a bottleneck for many important algorithms.
Any improvement in MM also leads
to more efficient algorithms for solving a plethora of other algebra
problems,
such as inverting matrices, solving systems of linear equations,
finding determinants, and also for some graph problems.

\subsection{Fixed Size Problems}

More or less all attempts to solve these problems
in the literature rely on solving,
once for all, certain fixed-size problems,
which can be the recursively applied to at lower levels,
to produce asymptotically fast algorithms.
\cite{Brent70,PanSpeed,Kaporin10K,IP6,BardApprox22}.

In 1969 Victor Strassen established a first asymptotic
improvement to the complexity of dense linear algebra,
by showing that two matrices 2x2 can be multiplied by using seven instead of
eight multiplications \cite{STR}.

Then in 1975 Laderman published a solution
for multiplying 3x3 matrices with 23 multiplications \cite{Laderman23}.
In 35 years this topic has generated 
very considerable interest, see for example 
\cite{PanSpeed,Kaporin10K,Kaporin2,SmithGrant,MakarovRussia22}
yet to this day it is not clear if Laderman's result 
is optimal and if it can be improved.



\subsection{Commutative Solutions}

Makarov found an 
algorithm using 22 multiplications 
for the product of 3x3 matrices 
but only in the commutative case, 
see \cite{MakarovRussia22}. 

\subsection{Approximate Solutions}

Very recently Gregory Bard found an approximate solution with 22
multiplications see \cite{BardApprox22}. 
An approximate solution with 21 was also found, see \cite{Approx21Reportedly}.
However it is much easier to find an approximate solution than an exact one.


\newpage

\section{New Result}

\subsection{Brent Equations}

As in many previous attempts to solve the problem we proceed by solving the so called Brent
equations \cite{Brent70}.
This approach has been tried many times before, see
\cite{Brent70,More23JohnsonMcLoughlin,SmithGrant,BardPhD,BardApprox22,BurrNicolaReportUCL}.

We write the coefficients of each products as three 3x3-matrices
for each multiplication
$A^{(i)}$, $B^{(i)}$ and $C^{(i)}$, $1 \le i \le r$, with $r=23$
where $A$ will be the left hand side of each product,
$B$ the right hand size, and $C$ tells to which coefficient of the
result this product contributes.

The Brent equations are as follows:

$$
\forall i \forall j \forall k \forall l \forall m \forall n~~
\sum_{i=1}^{r}
A^{(i)}_{ij} B^{(i)}_{kl} C^{(i)}_{mn} = \delta_{ni} \delta_{jk} \delta_{lm}
$$

For 3x3 matrixes we get 729 cubic equations exactly.

\subsection{Solving Brent Equations Modulo 2 and Lifting}

In general these equations can have
rational coefficients \cite{More23JohnsonMcLoughlin},
or even complex coefficients.
\cite{Approx21Reportedly}.

We are interested only in
very simple solutions which work over small finite
rings and fields.

First we write these Brent equations \cite{Brent70} modulo 2.
Then we solve them modulo 2.

Then we start from scratch and given a solution modulo 2,
we try to lift it by very similar formal
encoding and solving methods to a solution modulo 4.

So far a solution thus obtained seems to always be also a general solution
(for arbitrary rings $R$,
and therefore also for finite fields of arbitrary characteristic).
This is to say, we were quite lucky.

\subsection{Solving and Conversion}

Our equations are written algebraically and the converted to a SAT problem.
Our complete equations generator with some embedded converters to SAT
can be downloaded from \cite{HardProblemsPageCourtois}.

We have implemented a method to converting this very hard problem to a SAT problem,
and we have attempted to solve it,
with our portfolio of some 500 SAT solvers and their variants.
With many improvements and tweaks we are now able to obtain
such a solution for 23 variables in a few days with one single CPU.

\newpage
\subsection{The Laderman Solution From 1975}
\label{LadermanMethod23}

We present it in a form which can be directly verified with
Maple computer algebra software:

\begin{verbatim}
P01 := (a_1_1-a_1_2-a_1_3+a_2_1-a_2_2-a_3_2-a_3_3) *  (-b_2_2);
P02 := (a_1_1+a_2_1) *  (b_1_2+b_2_2);
P03 := (a_2_2) *  (b_1_1-b_1_2+b_2_1-b_2_2-b_2_3+b_3_1-b_3_3);
P04 := (-a_1_1-a_2_1+a_2_2) *  (-b_1_1+b_1_2+b_2_2);
P05 := (-a_2_1+a_2_2) *  (-b_1_1+b_1_2);
P06 := (a_1_1) *  (-b_1_1);
P07 := (a_1_1+a_3_1+a_3_2) *  (b_1_1-b_1_3+b_2_3);
P08 := (a_1_1+a_3_1) *  (-b_1_3+b_2_3);
P09 := (a_3_1+a_3_2) *  (b_1_1-b_1_3);
P10 := (a_1_1+a_1_2-a_1_3-a_2_2+a_2_3+a_3_1+a_3_2) *  (b_2_3);
P11 := (a_3_2) *  (-b_1_1+b_1_3+b_2_1-b_2_2-b_2_3-b_3_1+b_3_2);
P12 := (a_1_3+a_3_2+a_3_3) *  (b_2_2+b_3_1-b_3_2);
P13 := (a_1_3+a_3_3) *  (-b_2_2+b_3_2);
P14 := (a_1_3) *  (b_3_1);
P15 := (-a_3_2-a_3_3) *  (-b_3_1+b_3_2);
P16 := (a_1_3+a_2_2-a_2_3) *  (b_2_3-b_3_1+b_3_3);
P17 := (-a_1_3+a_2_3) *  (b_2_3+b_3_3);
P18 := (a_2_2-a_2_3) *  (b_3_1-b_3_3);
P19 := (a_1_2) *  (b_2_1);
P20 := (a_2_3) *  (b_3_2);
P21 := (a_2_1) *  (b_1_3);
P22 := (a_3_1) *  (b_1_2);
P23 := (a_3_3) *  (b_3_3);
expand(-P06+P14+P19-a_1_1*b_1_1-a_1_2*b_2_1-a_1_3*b_3_1);
expand(P01-P04+P05-P06-P12+P14+P15-a_1_1*b_1_2-a_1_2*b_2_2-a_1_3*b_3_2);
expand(-P06-P07+P09+P10+P14+P16+P18-a_1_1*b_1_3-a_1_2*b_2_3-a_1_3*b_3_3);
expand(P02+P03+P04+P06+P14+P16+P17-a_2_1*b_1_1-a_2_2*b_2_1-a_2_3*b_3_1);
expand(P02+P04-P05+P06+P20-a_2_1*b_1_2-a_2_2*b_2_2-a_2_3*b_3_2);
expand(P14+P16+P17+P18+P21-a_2_1*b_1_3-a_2_2*b_2_3-a_2_3*b_3_3);
expand(P06+P07-P08+P11+P12+P13-P14-a_3_1*b_1_1-a_3_2*b_2_1-a_3_3*b_3_1);
expand(P12+P13-P14-P15+P22-a_3_1*b_1_2-a_3_2*b_2_2-a_3_3*b_3_2);
expand(P06+P07-P08-P09+P23-a_3_1*b_1_3-a_3_2*b_2_3-a_3_3*b_3_3);
\end{verbatim}

\newpage
\subsection{The New Method with 23 Multiplications}
\label{NewMethodNotLaderman}

We present it in the same form which can also be directly verified
with Maple computer algebra software:

\begin{verbatim}
P01 := (a_2_3) *  (-b_1_2+b_1_3-b_3_2+b_3_3);
P02 := (-a_1_1+a_1_3+a_3_1+a_3_2) *  (b_2_1+b_2_2);
P03 := (a_1_3+a_2_3-a_3_3) *  (b_3_1+b_3_2-b_3_3);
P04 := (-a_1_1+a_1_3) *  (-b_2_1-b_2_2+b_3_1);
P05 := (a_1_1-a_1_3+a_3_3) *  (b_3_1);
P06 := (-a_2_1+a_2_3+a_3_1) *  (b_1_2-b_1_3);
P07 := (-a_3_1-a_3_2) *  (b_2_2);
P08 := (a_3_1) *  (b_1_1-b_2_1);
P09 := (-a_2_1-a_2_2+a_2_3) *  (b_3_3);
P10 := (a_1_1+a_2_1-a_3_1) *  (b_1_1+b_1_2+b_3_3);
P11 := (-a_1_2-a_2_2+a_3_2) *  (-b_2_2+b_2_3);
P12 := (a_3_3) *  (b_3_2);
P13 := (a_2_2) *  (b_1_3-b_2_3);
P14 := (a_2_1+a_2_2) *  (b_1_3+b_3_3);
P15 := (a_1_1) *  (-b_1_1+b_2_1-b_3_1);
P16 := (a_3_1) *  (b_1_2-b_2_2);
P17 := (a_1_2) *  (-b_2_2+b_2_3-b_3_3);
P18 := (-a_1_1+a_1_2+a_1_3+a_2_2+a_3_1) *  (b_2_1+b_2_2+b_3_3);
P19 := (-a_1_1+a_2_2+a_3_1) *  (b_1_3+b_2_1+b_3_3);
P20 := (-a_1_2+a_2_1+a_2_2-a_2_3-a_3_3) *  (-b_3_3);
P21 := (-a_2_2-a_3_1) *  (b_1_3-b_2_2);
P22 := (-a_1_1-a_1_2+a_3_1+a_3_2) *  (b_2_1);
P23 := (a_1_1+a_2_3) *  (b_1_2-b_1_3-b_3_1);
expand(P02+P04+P07-P15-P22-a_1_1*b_1_1-a_1_2*b_2_1-a_1_3*b_3_1);
expand(P01-P02+P03+P05-P07+P09+P12+P18-P19-P20-P21+P22+P23-
a_1_1*b_1_2-a_1_2*b_2_2-a_1_3*b_3_2);
expand(-P02-P07+P17+P18-P19-P21+P22-a_1_1*b_1_3-a_1_2*b_2_3-a_1_3*b_3_3);
expand(P06+P08+P10-P14+P15+P19-P23-a_2_1*b_1_1-a_2_2*b_2_1-a_2_3*b_3_1);
expand(-P01-P06+P09+P14+P16+P21-a_2_1*b_1_2-a_2_2*b_2_2-a_2_3*b_3_2);
expand(P09-P13+P14-a_2_1*b_1_3-a_2_2*b_2_3-a_2_3*b_3_3);
expand(P02+P04+P05+P07+P08-a_3_1*b_1_1-a_3_2*b_2_1-a_3_3*b_3_1);
expand(-P07+P12+P16-a_3_1*b_1_2-a_3_2*b_2_2-a_3_3*b_3_2);
expand(-P07-P09+P11-P13+P17+P20-P21-a_3_1*b_1_3-a_3_2*b_2_3-a_3_3*b_3_3);
\end{verbatim}

\newpage

\section{Equivalent Solutions}

An important question is as follows:
can our solution be obtained from
the Laderman's solution?

Equivalence relations
and the group of transformations
which allow to transform one
exact non-commutative solution for matrix multiplication,
into another such solution,
have been studied in
\cite{More23JohnsonMcLoughlin,DeGrooteGroupBilinear}.

We give here a brief description
of transformations in question:

As in \cite{More23JohnsonMcLoughlin}
we write the coefficients of each products as three 3x3-matrices
for each multiplication
$A^{(i)}$, $B^{(i)}$ and $C^{(i)}$, $1 \le i \le r$, with $r=23$
where $A$ will be the left hand side of each product,
$B$ the right hand size, and $C$ tells to which coefficient of the
result this product contributes.
%

We have the following transformations
which transform one solution to another solution:

\begin{enumerate}
\item
One can permute the $r$ indexes $i$.

\item
One can cyclically shift the three sets of matrices,
$A^{(i)}, B^{(i)}$ and $C^{(i)}$ for $1 \le i \le r$
becomes
$B^{(i)}, C^{(i)}$ and $A^{(i)}$ for $1 \le i \le r$.

\item
One reverse the order and transpose:
$A^{(i)}, B^{(i)}$ and $C^{(i)}$ for $1 \le i \le r$
becomes
$(C^{(i)})^T, (B^{(i)})^T$ and $(A^{(i)})^T$ for $1 \le i \le r$.

\item
One can rescale as follows:
$a_i A^{(i)}, b_i B^{(i)}$ and $c_i C^{(i)}$ for $1 \le i \le r$
where $a_i, b_i, c_i$ are rational coefficients with
$a_i b_i c_i=1$ for each  $1 \le i \le r$.

\item
This method is called "sandwiching".
We replace
$A^{(i)}, B^{(i)}$ and $C^{(i)}$ for $1 \le i \le r$
by
$U A^{(i)} V^{-1}$, $V B^{(i)} W^{-1}$ and $W C^{(i)} U^{-1}$, \\
where $U, V, W$ are three arbitrary invertible matrices.
\end{enumerate}

Main results in this area
of concern to us can be summarized as follows:
for the 2x2 case and 7 multiplications,
all non-commutative algorithms are equivalent to Strassen's algorithm,
see \cite{More23JohnsonMcLoughlin,DeGrooteGroupBilinear}.

For 3x3 matrices and 23 multiplications,
Johnson and McLoughlin have in 1986 exhibited
two families of infinitely
pairwise inequivalent algorithms,
see \cite{More23JohnsonMcLoughlin}.
Now the main question is,
is our solution new,
or already found in
\cite{More23JohnsonMcLoughlin}.

\newpage
\section{Comparison}

\subsection{Is Our Solution Equivalent To Any Previous Solution?}

An important question is as follows:
can our solution be obtained from
the Laderman's solution
or from one of the solutions from
\cite{More23JohnsonMcLoughlin}.

To prove inequivalence,
we follow the methodology of \cite{More23JohnsonMcLoughlin}.

\begin{theor}[Invariant for Equivalent Solutions]
\label{DefPermSol}
It is possible to see, that all the transformations
described on the previous page,
leave the distribution
of $3\times r$ ranks of matrices inchanged,
except that these integers can be permuted.
\end{theor}

\paragraph{Proof:}
\vskip-5pt
\vskip-5pt
This is obvious and
was already stated in
\cite{More23JohnsonMcLoughlin}.

\begin{theor}
\label{NotEquivToLaderman}
Our new solution
from Section \ref{NewMethodNotLaderman}
is neither equivalent
to the Laderman's solution
from Section
\ref{LadermanMethod23}
nor it is to any of the
solutions 
given in \cite{More23JohnsonMcLoughlin}.
\end{theor}
\vskip-5pt
\vskip-5pt

\paragraph{Proof:}
\vskip-5pt
\vskip-5pt
Following
\cite{More23JohnsonMcLoughlin},
the Laderman's solution
has exactly 6 matrices of rank 3
(which occur in products P01,P03,P06,P10,P11,P14 
in Section \ref{LadermanMethod23}).

At the same time in all new solutions presented in
\cite{More23JohnsonMcLoughlin},
at most 1 matrix will have rank 3.

In our solution we have exactly 2 matrices of rank 3
(which occur in products P18 and P20, 
they are 2 and not more such matrices, 
both being on the left hand size namely $A^{(18)}$, in $A^{(20)}$, 
and we have checked carefully, there is no mistake). 

This proves that all these solutions are distinct.



\newpage

\section{Conclusion}

One of the most famous problems in computer algebra
is the problem of fast matrix multiplication.
The progress in this area is very slow.
Many attempts to solve these problems
in the literature work by solving,
fixed-size problems and apply the solution recursively
\cite{Brent70,PanSpeed,Kaporin10K,IP6,BardApprox22}. 
This leads to pure combinatorial optimisation problems with fixed size.

In 1976 Laderman published a general and non-commutative
method to multiply two 3x3 matrices using only 23 multiplications.
In 35 years very little no progress was made on this very famous problem
and until this day
it remains an open problem if this can be done with 22 multiplications.

We have implemented a new method
which converts this very hard problem to a SAT problem,
and we have attempted to solve it,
with our portfolio of some 500 SAT solvers.
We were able to produce new solutions to the Laderman's problem.
We present a new
fully general and non-commutative solution
with also 23 multiplications.
We prove that this new solution
is {\bf NOT} equivalent to the Laderman's original solution,
neither it is equivalent to any of the new solutions given in \cite{More23JohnsonMcLoughlin}.
In fact it is very different.

This preliminary result gives strong evidence that the space of solutions
to Laderman's problem is larger than expected,
and therefore it is worth trying to find more such solutions.
It further increases the chances that a solution for 22 multiplications exits and
it might be found soon by running our algorithms longer,
or just by using better SAT solvers.
This also motivates further research
about SAT solvers and their applications
in mathematics and computer science.

\paragraph{\bf Acknowledgements:~}
We would like to thank
Richard Brent, Markus Bl\"aser, 
Denis Farr and Alexey Pospelov 
for their very helpful comments. 

\vfill
\pagebreak


\begin{thebibliography}{99}

    \bibitem{BlaserLowerBounds}
Markus Bl\"aser:
\sl On the complexity of the multiplication of matrices of small formats, \rm
In Journal of Complexity 19(1): 43-60 (2003).


%
    \bibitem{BardApprox22}
Gregory Bard:
\sl New Practical Approximate Matrix Multiplication Algorithms found via Solving a System of Cubic Equations.
\rm A draft paper submitted to a journal, can be found at:
\url{http://www-users.math.umd.edu/{\textasciitilde}bardg/}

    \bibitem{BardPhD} Gregory Bard:
\sl Algorithms for Solving Linear and Polynomial Systems of
Equations over Finite Fields with Applications to Cryptanalysis, \rm
Submitted in Partial Fulfillment for the degree of Doctor of Philosophy of
Applied Mathematics and Scientific Computation.
PhD Thesis, University of Maryland at College Park, April 30, 2007.

    \bibitem{PeraltaMultComp}
  	Joan Boyar, Ren\'{e} Peralta:
  \sl A New Combinational Logic Minimization Technique with Applications to Cryptology. \rm
In SEA 2010: 178-189. \\
An early version was published in 2009 at
\url{http://eprint.iacr.org/2009/191}. It was revised 13 Mar 2010.


    \bibitem{PeraltaWeb}
Michael Bartock, Joan Boyar, Morris Dworkin, Michael Fischer, Rene Peralta,
Bruce Strackbein, Catie Baker, Andrea Visconti, Chiara Schiavo, Johnny Svensson,
Holman Gao, Scott Zimmermann, Matteo Bocchi:
\sl
Circuit Minimization Work, \rm
A web page which gives the solutions to various optimisations produced
by the CMT team at the University of Yale, USA.
\url{http://cs-www.cs.yale.edu/homes/peralta/CircuitStuff/CMT.html}.
%

    \bibitem{Brent70}
Richard Brent:
\sl Algorithms for matrix multiplication, \rm
Tech. Report Report TR-CS-70-157,
Department of Computer Science, Stanford, 52 pages, March 1970.
Available at
\url{http://maths.anu.edu.au/~brent/pd/rpb002i.pdf}

    \bibitem{BurrNicolaReportUCL}
Nicola Burr:
\sl
An investigation into fast matrix multiplication, \rm
individual research project report,
done under supervision of Nicolas T. Courtois,
and submitted as a part of BSc Degree in Computer Science
at Univesity College London,
6 June 2010.

    \bibitem{KohnGroupBreaktroughFastMM}
Henry Cohn, Robert Kleinberg, Balazs Szegedyz and Christopher Umans:
\sl Group-theoretic Algorithms for Matrix Multiplication, \rm
In FOCS'05,
46th Annual IEEE Symposium on Foundations of Computer Science,
pp. 379.

     \bibitem{CPW} Don Coppersmith, Shmuel Winograd:
\sl Matrix multiplication via arithmetic progressions, \rm J.
Symbolic Computation (1990), 9, pp. 251-280.

    \bibitem{CPWAsy} Don Coppersmith, Shmuel Winograd,
{\sl On the asymptotic complexity of matrix multiplication},
SIAM Journal Comp., 11(1980), pp 472-492.


    \bibitem{HardProblemsPageCourtois}
Nicolas T. Courtois,
\sl
Benchmarking Algebraic,
Logical and Constraint Solvers
and Study of Selected Hard Problems, \rm
Web page with software,
see last section entitled:
Multiplicative Complexity Challenges - Linear Algebra,
at \url{http://www.cryptosystem.net/aes/hardproblems.html}.

    \bibitem{CourtoisSoftware}
    Nicolas Courtois:
Some algebraic cryptanalysis software made available for free,
by Nicolas T. Courtois,
\url{http://www.cryptosystem.net/aes/tools.html}.


    \bibitem{DeGrooteGroupBilinear}
Hans F. de Groote:
\sl On Varieties of Optimal Algorithms for the Computation of Bilinear Mappings I.
The Isotropy Group of a Bilinear Mapping, \rm
In Theor. Comput. Sci. 7, pp. 1-24, 1978.

    \bibitem{gus} John Gustafson, Srinivas Aluru,
\sl Massively Parallel Searching for Better Algorithms
or, How to Do a Cross Product with Five Multiplications, \rm
Ames Laboratory, Department of Energy, ISU, Ames, Iowa.
Available at \url{http://www.scl.ameslab.gov/Publications/FiveMultiplications/Five.html}

    \bibitem{has} Johan H{\aa}stad,
{\sl Tensor Rank is NP-Complete\/},
Journal of Algorithms, vol. 11, pp. 644-654, 1990.

    \bibitem{More23JohnsonMcLoughlin}
Rodney W. Johnson and Aileen M. McLoughlin:
\sl Noncommutative Bilinear Algorithms for 3 x 3 Matrix Multiplication, \rm
In SIAM J. Comput., vol. 15 (2), pp. 595-603, 1986.

    \bibitem{Kaporin10K}
Igor Kaporin:
\sl A practical algorithm for faster matrix multiplication, \rm
In Numerical Linear Algebra with Applications Volume 6, Issue 8, pages 687–700, December 1999.
\url{http://atlas.mat.ub.es/personals/sombra/cours/uba/programas/kaporin99.pdf}

    \bibitem{Kaporin2}
Igor Kaporin:
\sl The aggregation and cancellation techniques as a practical tool for faster matrix multiplication, \rm
In Journal Theoretical Computer Science - Algebraic and numerical algorithm archive
Volume 315 Issue 2-3, 6 May 2004.

    \bibitem{Laderman23}
Julian D. Laderman:
\sl
A Non-Commutative Algorithm for Multiplying 3x3 Matrices Using 23 Multiplications, \rm
Bull. Amer. Math. Soc. Volume 82, Number 1 (1976), 126-128.

    \bibitem{MakarovRussia22}
    O. M. Makarov:
\sl An algorithm for multiplication of 3 x 3 matrices. \rm
Zh. Vychisl. Mat. i Mat. Fiz., 26(2):293–294, 320, 1986.



    \bibitem{IP6}  Jacques Patarin, Nicolas Courtois , Louis Goubin:
{\sl Improved Algorithms for Isomorphism of Polynomials}; Eurocrypt 1998, Springer.



\bibitem{PanSpeed}
Victor Pan,
\sl How Can We Speed Up Matrix Multiplication?
In  SIAM Review, Volume 26 Issue 3, pp 393-415, July 1984.

\bibitem{RobinsonSIAM}
Sara Robinson:
\sl Toward an Optimal Algorithm for Matrix Multiplication,
\rm
In SIAM News, vol. 38 number 9, November 2005.

    \bibitem{Approx21Reportedly}
Arnold Sch\"onhage:
\sl Partial and Total Matrix Multiplication, \rm
SIAM J. Comput. vol. 10 (3) pp. 434-455, 1981.

    \bibitem{SmithGrant}
Warren Smith:
\sl
Fast Matrix Algorithms And Multiplication Formulae, \rm
Available at \url{https://math.cst.temple.edu/~wds/matgrant.ps}.

    \bibitem{Sykora}
Ondrej S\'ykora:
\sl A fast non-commutative algorithm for matrix multiplication, \rm
In Mathematical foundations of computer science
(Proc. Sixth Sympos., Tatransk\'{a} Lomnica, 1977),
Springer, 1977, p. 504–512. LNCS 53.

    \bibitem{STR} Volker Strassen, {\sl Gaussian elimination is not optimal\/},
Numerische Mathematik 13, 1969, pp. 354-356.


\end{thebibliography}
\end{document}